\documentclass{PoS}

\usepackage{amsmath,amssymb}
\usepackage{graphicx,graphics}
\usepackage{xspace}

\title{QCD and Electroweak Physics at LHC}
\ShortTitle{QCD and Electroweak Physics at LHC}
\author{\speaker{Klaus Rabbertz}\\
  on behalf of the ATLAS and CMS Collaborations\\
  Institut f\"ur Experimentelle Kernphysik, KIT --- Karlsruher
  Institut f\"ur Technologie, Germany\\
  E-mail: \email{klaus.rabbertz@cern.ch}}
\abstract{%
  First LHC data have been collected and collisions at a
  center-of-mass energy of $7\,$TeV are anticipated for the next
  months.  The commissioning of the detectors and the re-establishment
  of the Standard Model in the new energy regime will be the main
  tasks for the experimental collaborations in the year to come.  This
  report summarizes the measurement plans and performance expectations
  of the ATLAS and CMS experiments for a selected number of QCD and
  electroweak analyses with an emphasis on the early data taking
  phase. Some longer term prospects are pointed out.%
}
\FullConference{RADCOR 2009 - 9th International Symposium on Radiative Corrections (Applications of Quantum Field Theory to Phenomenology) \\
  October 25-30 2009\\
  Ascona, Switzerland}
\begin{document}
%
%
%
%
%
%

\newcommand {\etal}{\mbox{et al.}\xspace} 
\newcommand {\ie}{\mbox{i.e.}\xspace}     
\newcommand {\eg}{\mbox{e.g.}\xspace}     
\newcommand {\etc}{\mbox{etc.}\xspace}     
\newcommand {\vs}{\mbox{\sl vs.}\xspace}      
\newcommand {\mdash}{\ensuremath{\mathrm{-}}} 

\newcommand {\Lone}{Level-1\xspace} 
\newcommand {\Ltwo}{Level-2\xspace}
\newcommand {\Lthree}{Level-3\xspace}

\providecommand{\ACERMC} {\textsc{AcerMC}\xspace}
\providecommand{\ALPGEN} {{\textsc{alpgen}}\xspace}
\providecommand{\CHARYBDIS} {{\textsc{charybdis}}\xspace}
\providecommand{\CMKIN} {\textsc{cmkin}\xspace}
\providecommand{\CMSIM} {{\textsc{cmsim}}\xspace}
\providecommand{\CMSSW} {{\textsc{cmssw}}\xspace}
\providecommand{\COBRA} {{\textsc{cobra}}\xspace}
\providecommand{\COCOA} {{\textsc{cocoa}}\xspace}
\providecommand{\COMPHEP} {\textsc{CompHEP}\xspace}
\providecommand{\EVTGEN} {{\textsc{evtgen}}\xspace}
\providecommand{\FAMOS} {{\textsc{famos}}\xspace}
\providecommand{\GARCON} {\textsc{garcon}\xspace}
\providecommand{\GARFIELD} {{\textsc{garfield}}\xspace}
\providecommand{\GEANE} {{\textsc{geane}}\xspace}
\providecommand{\GEANTfour} {{\textsc{geant4}}\xspace}
\providecommand{\GEANTthree} {{\textsc{geant3}}\xspace}
\providecommand{\GEANT} {{\textsc{geant}}\xspace}
\providecommand{\HDECAY} {\textsc{hdecay}\xspace}
\providecommand{\HERWIG} {{\textsc{herwig}}\xspace}
\providecommand{\HIGLU} {{\textsc{higlu}}\xspace}
\providecommand{\HIJING} {{\textsc{hijing}}\xspace}
\providecommand{\IGUANA} {\textsc{iguana}\xspace}
\providecommand{\ISAJET} {{\textsc{isajet}}\xspace}
\providecommand{\ISAPYTHIA} {{\textsc{isapythia}}\xspace}
\providecommand{\ISASUGRA} {{\textsc{isasugra}}\xspace}
\providecommand{\ISASUSY} {{\textsc{isasusy}}\xspace}
\providecommand{\ISAWIG} {{\textsc{isawig}}\xspace}
\providecommand{\MADGRAPH} {\textsc{MadGraph}\xspace}
\providecommand{\MCATNLO} {\textsc{mc@nlo}\xspace}
\providecommand{\MCFM} {\textsc{mcfm}\xspace}
\providecommand{\MILLEPEDE} {{\textsc{millepede}}\xspace}
\providecommand{\ORCA} {{\textsc{orca}}\xspace}
\providecommand{\OSCAR} {{\textsc{oscar}}\xspace}
\providecommand{\PHOTOS} {\textsc{photos}\xspace}
\providecommand{\PROSPINO} {\textsc{prospino}\xspace}
\providecommand{\PYTHIA} {{\textsc{pythia}}\xspace}
\providecommand{\SHERPA} {{\textsc{sherpa}}\xspace}
\providecommand{\TAUOLA} {\textsc{tauola}\xspace}
\providecommand{\TOPREX} {\textsc{TopReX}\xspace}
\providecommand{\XDAQ} {{\textsc{xdaq}}\xspace}

\newcommand {\DZERO}{D\O\xspace}     


\newcommand{\de}{\ensuremath{^\circ}}
\newcommand{\ten}[1]{\ensuremath{\times \text{10}^\text{#1}}}
\newcommand{\unit}[1]{\ensuremath{\text{\,#1}}\xspace}
\newcommand{\mum}{\ensuremath{\,\mu\text{m}}\xspace}
\newcommand{\micron}{\ensuremath{\,\mu\text{m}}\xspace}
\newcommand{\cm}{\ensuremath{\,\text{cm}}\xspace}
\newcommand{\mm}{\ensuremath{\,\text{mm}}\xspace}
\newcommand{\mus}{\ensuremath{\,\mu\text{s}}\xspace}
\newcommand{\keV}{\ensuremath{\,\text{ke\hspace{-.08em}V}}\xspace}
\newcommand{\MeV}{\ensuremath{\,\text{Me\hspace{-.08em}V}}\xspace}
\newcommand{\GeV}{\ensuremath{\,\text{Ge\hspace{-.08em}V}}\xspace}
\newcommand{\TeV}{\ensuremath{\,\text{Te\hspace{-.08em}V}}\xspace}
\newcommand{\PeV}{\ensuremath{\,\text{Pe\hspace{-.08em}V}}\xspace}
\newcommand{\keVc}{\ensuremath{{\,\text{ke\hspace{-.08em}V\hspace{-0.16em}/\hspace{-0.08em}c}}}\xspace}
\newcommand{\MeVc}{\ensuremath{{\,\text{Me\hspace{-.08em}V\hspace{-0.16em}/\hspace{-0.08em}c}}}\xspace}
\newcommand{\GeVc}{\ensuremath{{\,\text{Ge\hspace{-.08em}V\hspace{-0.16em}/\hspace{-0.08em}c}}}\xspace}
\newcommand{\TeVc}{\ensuremath{{\,\text{Te\hspace{-.08em}V\hspace{-0.16em}/\hspace{-0.08em}c}}}\xspace}
\newcommand{\keVcc}{\ensuremath{{\,\text{ke\hspace{-.08em}V\hspace{-0.16em}/\hspace{-0.08em}c}^\text{2}}}\xspace}
\newcommand{\MeVcc}{\ensuremath{{\,\text{Me\hspace{-.08em}V\hspace{-0.16em}/\hspace{-0.08em}c}^\text{2}}}\xspace}
\newcommand{\GeVcc}{\ensuremath{{\,\text{Ge\hspace{-.08em}V\hspace{-0.16em}/\hspace{-0.08em}c}^\text{2}}}\xspace}
\newcommand{\TeVcc}{\ensuremath{{\,\text{Te\hspace{-.08em}V\hspace{-0.16em}/\hspace{-0.08em}c}^\text{2}}}\xspace}

\newcommand{\pbinv} {\mbox{\ensuremath{\,\text{pb}^\text{$-$1}}}\xspace}
\newcommand{\fbinv} {\mbox{\ensuremath{\,\text{fb}^\text{$-$1}}}\xspace}
\newcommand{\nbinv} {\mbox{\ensuremath{\,\text{nb}^\text{$-$1}}}\xspace}
\newcommand{\percms}{\ensuremath{\,\text{cm}^\text{$-$2}\,\text{s}^\text{$-$1}}\xspace}
\newcommand{\lumi}{\ensuremath{\mathcal{L}}\xspace}
\newcommand{\Lumi}{\ensuremath{\mathcal{L}}\xspace}
%
\newcommand{\LvLow}  {\ensuremath{\mathcal{L}=\text{10}^\text{32}\,\text{cm}^\text{$-$2}\,\text{s}^\text{$-$1}}\xspace}
\newcommand{\LLow}   {\ensuremath{\mathcal{L}=\text{10}^\text{33}\,\text{cm}^\text{$-$2}\,\text{s}^\text{$-$1}}\xspace}
\newcommand{\lowlumi}{\ensuremath{\mathcal{L}=\text{2}\times \text{10}^\text{33}\,\text{cm}^\text{$-$2}\,\text{s}^\text{$-$1}}\xspace}
\newcommand{\LMed}   {\ensuremath{\mathcal{L}=\text{2}\times \text{10}^\text{33}\,\text{cm}^\text{$-$2}\,\text{s}^\text{$-$1}}\xspace}
\newcommand{\LHigh}  {\ensuremath{\mathcal{L}=\text{10}^\text{34}\,\text{cm}^\text{-2}\,\text{s}^\text{$-$1}}\xspace}
\newcommand{\hilumi} {\ensuremath{\mathcal{L}=\text{10}^\text{34}\,\text{cm}^\text{-2}\,\text{s}^\text{$-$1}}\xspace}


\newcommand{\zp}{\ensuremath{\mathrm{Z}^\prime}\xspace}


\newcommand{\kt}{\ensuremath{k_{\mathrm{T}}}\xspace}
\newcommand{\BC}{\ensuremath{{B_{\mathrm{c}}}}\xspace}
\newcommand{\bbarc}{\ensuremath{{\overline{b}c}}\xspace}
\newcommand{\bbbar}{\ensuremath{{b\overline{b}}}\xspace}
\newcommand{\ccbar}{\ensuremath{{c\overline{c}}}\xspace}
\newcommand{\JPsi}{\ensuremath{{J}/\psi}\xspace}
\newcommand{\bspsiphi}{\ensuremath{B_s \to \JPsi\, \phi}\xspace}
\newcommand{\AFB}{\ensuremath{A_\mathrm{FB}}\xspace}
\newcommand{\EE}{\ensuremath{e^+e^-}\xspace}
\newcommand{\MM}{\ensuremath{\mu^+\mu^-}\xspace}
\newcommand{\TT}{\ensuremath{\tau^+\tau^-}\xspace}
\newcommand{\wangle}{\ensuremath{\sin^{2}\theta_{\mathrm{eff}}^\mathrm{lept}(M^2_\mathrm{Z})}\xspace}
\newcommand{\ttbar}{\ensuremath{{t\overline{t}}}\xspace}
\newcommand{\stat}{\ensuremath{\,\text{(stat.)}}\xspace}
\newcommand{\syst}{\ensuremath{\,\text{(syst.)}}\xspace}

\newcommand{\HGG}{\ensuremath{\mathrm{H}\to\gamma\gamma}}
\newcommand{\gev}{\GeV}
\newcommand{\GAMJET}{\ensuremath{\gamma + \mathrm{jet}}}
\newcommand{\PPTOJETS}{\ensuremath{\mathrm{pp}\to\mathrm{jets}}}
\newcommand{\PPTOGG}{\ensuremath{\mathrm{pp}\to\gamma\gamma}}
\newcommand{\PPTOGAMJET}{\ensuremath{\mathrm{pp}\to\gamma +
\mathrm{jet}
}}
\newcommand{\MH}{\ensuremath{\mathrm{M_{\mathrm{H}}}}}
\newcommand{\RNINE}{\ensuremath{\mathrm{R}_\mathrm{9}}}
\newcommand{\DR}{\ensuremath{\Delta\mathrm{R}}}


\newcommand{\PT}{\ensuremath{p_{\mathrm{T}}}\xspace}
\newcommand{\pt}{\ensuremath{p_{\mathrm{T}}}\xspace}
\newcommand{\ET}{\ensuremath{E_{\mathrm{T}}}\xspace}
\newcommand{\HT}{\ensuremath{H_{\mathrm{T}}}\xspace}
\newcommand{\et}{\ensuremath{E_{\mathrm{T}}}\xspace}
\newcommand{\Em}{\ensuremath{E\!\!\!/}\xspace}
\newcommand{\Pm}{\ensuremath{p\!\!\!/}\xspace}
\newcommand{\PTm}{\ensuremath{{p\!\!\!/}_{\mathrm{T}}}\xspace}
\newcommand{\ETm}{\ensuremath{E_{\mathrm{T}}^{\mathrm{miss}}}\xspace}
\newcommand{\MET}{\ensuremath{E_{\mathrm{T}}^{\mathrm{miss}}}\xspace}
\newcommand{\ETmiss}{\ensuremath{E_{\mathrm{T}}^{\mathrm{miss}}}\xspace}
\newcommand{\VEtmiss}{\ensuremath{{\vec E}_{\mathrm{T}}^{\mathrm{miss}}}\xspace}

%

\newcommand{\ga}{\ensuremath{\gtrsim}}
\newcommand{\la}{\ensuremath{\lesssim}}
\newcommand{\swsq}{\ensuremath{\sin^2\theta_W}\xspace}
\newcommand{\cwsq}{\ensuremath{\cos^2\theta_W}\xspace}
\newcommand{\tanb}{\ensuremath{\tan\beta}\xspace}
\newcommand{\tanbsq}{\ensuremath{\tan^{2}\beta}\xspace}
\newcommand{\sidb}{\ensuremath{\sin 2\beta}\xspace}
\newcommand{\alpS}{\ensuremath{\alpha_S}\xspace}
\newcommand{\alpt}{\ensuremath{\tilde{\alpha}}\xspace}

\newcommand{\QL}{\ensuremath{Q_L}\xspace}
\newcommand{\sQ}{\ensuremath{\tilde{Q}}\xspace}
\newcommand{\sQL}{\ensuremath{\tilde{Q}_L}\xspace}
\newcommand{\ULC}{\ensuremath{U_L^C}\xspace}
\newcommand{\sUC}{\ensuremath{\tilde{U}^C}\xspace}
\newcommand{\sULC}{\ensuremath{\tilde{U}_L^C}\xspace}
\newcommand{\DLC}{\ensuremath{D_L^C}\xspace}
\newcommand{\sDC}{\ensuremath{\tilde{D}^C}\xspace}
\newcommand{\sDLC}{\ensuremath{\tilde{D}_L^C}\xspace}
\newcommand{\LL}{\ensuremath{L_L}\xspace}
\newcommand{\sL}{\ensuremath{\tilde{L}}\xspace}
\newcommand{\sLL}{\ensuremath{\tilde{L}_L}\xspace}
\newcommand{\ELC}{\ensuremath{E_L^C}\xspace}
\newcommand{\sEC}{\ensuremath{\tilde{E}^C}\xspace}
\newcommand{\sELC}{\ensuremath{\tilde{E}_L^C}\xspace}
\newcommand{\sEL}{\ensuremath{\tilde{E}_L}\xspace}
\newcommand{\sER}{\ensuremath{\tilde{E}_R}\xspace}
\newcommand{\sFer}{\ensuremath{\tilde{f}}\xspace}
\newcommand{\sQua}{\ensuremath{\tilde{q}}\xspace}
\newcommand{\sUp}{\ensuremath{\tilde{u}}\xspace}
\newcommand{\suL}{\ensuremath{\tilde{u}_L}\xspace}
\newcommand{\suR}{\ensuremath{\tilde{u}_R}\xspace}
\newcommand{\sDw}{\ensuremath{\tilde{d}}\xspace}
\newcommand{\sdL}{\ensuremath{\tilde{d}_L}\xspace}
\newcommand{\sdR}{\ensuremath{\tilde{d}_R}\xspace}
\newcommand{\sTop}{\ensuremath{\tilde{t}}\xspace}
\newcommand{\stL}{\ensuremath{\tilde{t}_L}\xspace}
\newcommand{\stR}{\ensuremath{\tilde{t}_R}\xspace}
\newcommand{\stone}{\ensuremath{\tilde{t}_1}\xspace}
\newcommand{\sttwo}{\ensuremath{\tilde{t}_2}\xspace}
\newcommand{\sBot}{\ensuremath{\tilde{b}}\xspace}
\newcommand{\sbL}{\ensuremath{\tilde{b}_L}\xspace}
\newcommand{\sbR}{\ensuremath{\tilde{b}_R}\xspace}
\newcommand{\sbone}{\ensuremath{\tilde{b}_1}\xspace}
\newcommand{\sbtwo}{\ensuremath{\tilde{b}_2}\xspace}
\newcommand{\sLep}{\ensuremath{\tilde{l}}\xspace}
\newcommand{\sLepC}{\ensuremath{\tilde{l}^C}\xspace}
\newcommand{\sEl}{\ensuremath{\tilde{e}}\xspace}
\newcommand{\sElC}{\ensuremath{\tilde{e}^C}\xspace}
\newcommand{\seL}{\ensuremath{\tilde{e}_L}\xspace}
\newcommand{\seR}{\ensuremath{\tilde{e}_R}\xspace}
\newcommand{\snL}{\ensuremath{\tilde{\nu}_L}\xspace}
\newcommand{\sMu}{\ensuremath{\tilde{\mu}}\xspace}
\newcommand{\sNu}{\ensuremath{\tilde{\nu}}\xspace}
\newcommand{\sTau}{\ensuremath{\tilde{\tau}}\xspace}
\newcommand{\Glu}{\ensuremath{g}\xspace}
\newcommand{\sGlu}{\ensuremath{\tilde{g}}\xspace}
\newcommand{\Wpm}{\ensuremath{W^{\pm}}\xspace}
\newcommand{\sWpm}{\ensuremath{\tilde{W}^{\pm}}\xspace}
\newcommand{\Wz}{\ensuremath{W^{0}}\xspace}
\newcommand{\sWz}{\ensuremath{\tilde{W}^{0}}\xspace}
\newcommand{\sWino}{\ensuremath{\tilde{W}}\xspace}
\newcommand{\Bz}{\ensuremath{B^{0}}\xspace}
\newcommand{\sBz}{\ensuremath{\tilde{B}^{0}}\xspace}
\newcommand{\sBino}{\ensuremath{\tilde{B}}\xspace}
\newcommand{\Zz}{\ensuremath{Z^{0}}\xspace}
\newcommand{\sZino}{\ensuremath{\tilde{Z}^{0}}\xspace}
\newcommand{\sGam}{\ensuremath{\tilde{\gamma}}\xspace}
\newcommand{\chiz}{\ensuremath{\tilde{\chi}^{0}}\xspace}
\newcommand{\chip}{\ensuremath{\tilde{\chi}^{+}}\xspace}
\newcommand{\chim}{\ensuremath{\tilde{\chi}^{-}}\xspace}
\newcommand{\chipm}{\ensuremath{\tilde{\chi}^{\pm}}\xspace}
\newcommand{\Hone}{\ensuremath{H_{d}}\xspace}
\newcommand{\sHone}{\ensuremath{\tilde{H}_{d}}\xspace}
\newcommand{\Htwo}{\ensuremath{H_{u}}\xspace}
\newcommand{\sHtwo}{\ensuremath{\tilde{H}_{u}}\xspace}
\newcommand{\sHig}{\ensuremath{\tilde{H}}\xspace}
\newcommand{\sHa}{\ensuremath{\tilde{H}_{a}}\xspace}
\newcommand{\sHb}{\ensuremath{\tilde{H}_{b}}\xspace}
\newcommand{\sHpm}{\ensuremath{\tilde{H}^{\pm}}\xspace}
\newcommand{\hz}{\ensuremath{h^{0}}\xspace}
\newcommand{\Hz}{\ensuremath{H^{0}}\xspace}
\newcommand{\Az}{\ensuremath{A^{0}}\xspace}
\newcommand{\Hpm}{\ensuremath{H^{\pm}}\xspace}
\newcommand{\sGra}{\ensuremath{\tilde{G}}\xspace}
\newcommand{\mtil}{\ensuremath{\tilde{m}}\xspace}
\newcommand{\rpv}{\ensuremath{\rlap{\kern.2em/}R}\xspace}
\newcommand{\LLE}{\ensuremath{LL\bar{E}}\xspace}
\newcommand{\LQD}{\ensuremath{LQ\bar{D}}\xspace}
\newcommand{\UDD}{\ensuremath{\overline{UDD}}\xspace}
\newcommand{\Lam}{\ensuremath{\lambda}\xspace}
\newcommand{\Lamp}{\ensuremath{\lambda'}\xspace}
\newcommand{\Lampp}{\ensuremath{\lambda''}\xspace}
\newcommand{\spinbd}[2]{\ensuremath{\bar{#1}_{\dot{#2}}}\xspace}

\newcommand{\MD}{\ensuremath{{M_\mathrm{D}}}\xspace}
\newcommand{\Mpl}{\ensuremath{{M_\mathrm{Pl}}}\xspace}
\newcommand{\Rinv} {\ensuremath{{R}^{-1}}\xspace}

%
%
\hyphenation{en-viron-men-tal}

%
%
\DeclareGraphicsExtensions{.pdf,.png}
\graphicspath{{pdf/},{png/}} 

%
%
\providecommand{\FIXME}[1]{({\bf FIXME: #1})}
\providecommand{\HERWIGPP} {{\textsc{herwig++}}\xspace}
\providecommand{\NLOJETPP} {{\textsc{nlojet++}}\xspace}
\providecommand{\PHOJET} {{\textsc{phojet}}\xspace}
\providecommand{\Et}{E_{\mathrm{T}}}
\providecommand{\met}{\mbox{${\hbox{$\vec{E}$\kern-0.5em\lower-.1ex\hbox{/}}}_T~$}}
\providecommand{\MET}{\mbox{${\hbox{$E$\kern-0.5em\lower-.1ex\hbox{/}}}_T~$}}
\providecommand{\pthat}{\ensuremath{\hat{\text{p}}_\mathrm{T}}\xspace}
\providecommand{\kthat}{\ensuremath{\hat{\text{k}}_\mathrm{T}}\xspace}
\providecommand{\xt}{\ensuremath{\text{x}_\mathrm{T}}\xspace}
\providecommand{\ptjet}{\ensuremath{p_{\mathrm{T,jet}}}\xspace}
\providecommand{\FASTNLO} {{fast\textsc{nlo}}\xspace}

\providecommand{\order}{{\cal O}}
\providecommand{\rbthm}{\rule[-2ex]{0ex}{5ex}}
\providecommand{\rbthr}{\rule[-1.7ex]{0ex}{5ex}}
\providecommand{\rbtrm}{\rule[-2ex]{0ex}{5ex}}
\providecommand{\rbtrr}{\rule[-0.8ex]{0ex}{3.2ex}}
\providecommand{\relmet}{\ensuremath{\text{MET}/\sum{E_{\mathrm{T}}}}\xspace}
\providecommand{\LvStartup}{\ensuremath{\mathcal{L}=\text{10}^\text{31}\,\text{cm}^\text{$-$2}\,\text{s}^\text{$-$1}}\xspace}

%
%
%
%

\section{Introduction}

In November 2009 the Large Hadron Collider (LHC) at CERN has been
restarted and first collisions at the injection energy of $\sqrt{s} =
900\GeV$ have been registered. On the 13th of December the proton
beams could even be accelerated up to $1.18\TeV$ and achieved a record
collision energy of $2.36\TeV$ dethroning the Tevatron at Fermilab as
the most powerful particle accelerator in the world.  A new era in
particle physics has just started and even higher center-of-mass
energies up to $7\TeV$ are anticipated for the next months.

Reaching, however, the high level of understanding of the complex
detectors and the accelerator that is required in order to perform
measurements as accurate as reported here~\cite{Wallny:RADCOR2009}
from the Tevatron experiments CDF and D0 will take time.  The
commissioning of the detectors and the re-establishment of the
Standard Model in the new energy regime will therefore be the main
tasks for the experimental collaborations in the year to come.  This
report summarizes the measurement plans and performance expectations
of the ATLAS and CMS experiments for a selected number of QCD and
electroweak analyses with an emphasis on this early data taking phase.
Some longer term prospects are pointed out.  Detailed descriptions of
the LHC and the four main experiments can be found
elsewhere~\cite{Evans:2008zzb,Aamodt:2008zz,Aad:2008zzm,CMS-DET,Alves:2008zz}.

\section{Minimum Bias and Underlying Event}

\begin{figure}[p]
  \centering
  \includegraphics[width=0.45\textwidth]{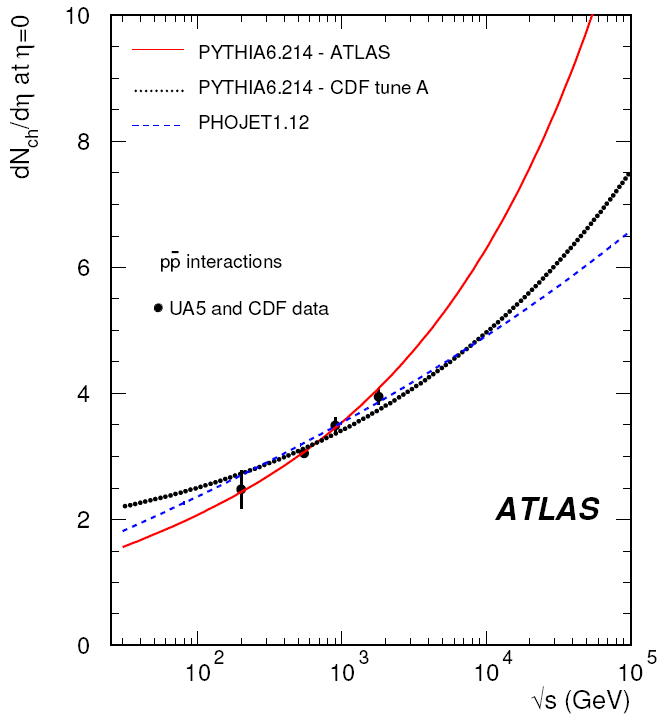}%
  \includegraphics[width=0.55\textwidth]{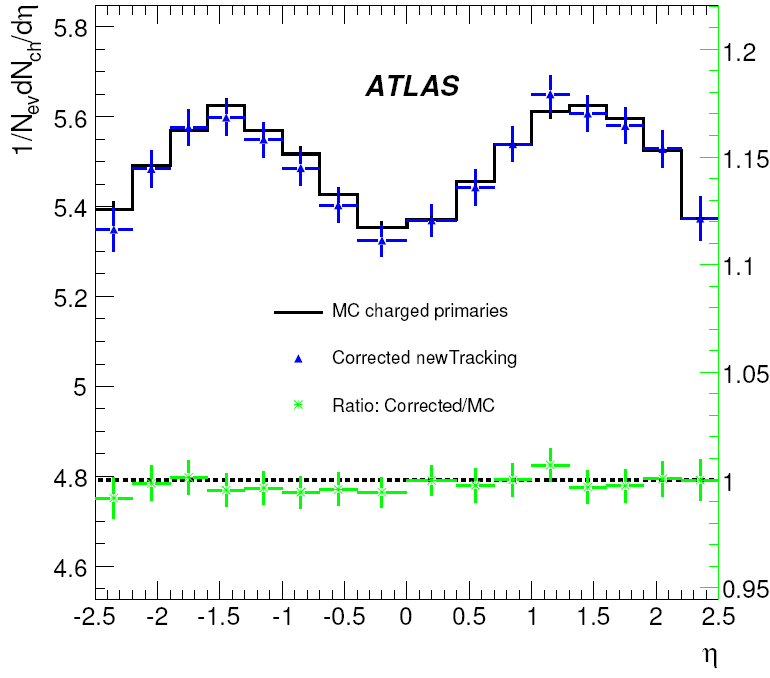}
  \caption{The average charged-particle density at central rapidity as
    a function of the center-of-mass energy is deduced from
    simulations with different tunes of the MC models \PYTHIA and
    \PHOJET on the left~\cite{Aad:2009wy}. An example of the
    reconstruction of the charged-particle density using the ATLAS
    tracking for $|\eta| < 2.5$ at $14\TeV$ is given on the
    right~\cite{Aad:2009wy}.}
  \label{fig:dNdetavsEcms}
\end{figure}

Charged-particle multiplicity distributions versus pseudorapidity
$\eta=-\ln\tan(\theta/2)$ (with $\theta$ being the polar angle) and
transverse momentum $p_T$ have a long tradition in hadron-hadron
collisions and have been measured e.g.\ in experiments at the ISR,
SPS, Tevatron and RHIC
colliders~\cite{Thome:1977ky,Alner:1986xu,Abe:1989td,Nouicer:2004ke}.
Since in terms of theory only models are available, the predictions
for higher energies vary significantly even though they have been
tuned to describe currently available data.  The expectations by the
ATLAS Collaboration~\cite{Aad:2009wy} from
\PYTHIA~\cite{Sjostrand:2006za} and \PHOJET~\cite{Bopp:1998rc} for the
average charged-particle density at central rapidity versus the
center-of-mass energy is presented in Figure~\ref{fig:dNdetavsEcms}
left. An example of the reconstruction of the charged-particle density
$dN_{ch}/d\eta$ using the ATLAS tracking for $|\eta| < 2.5$ at
$14\TeV$ is given in Figure~\ref{fig:dNdetavsEcms}
right~\cite{Aad:2009wy}.

\begin{figure}[p]
  \centering
  \includegraphics[width=0.50\textwidth]{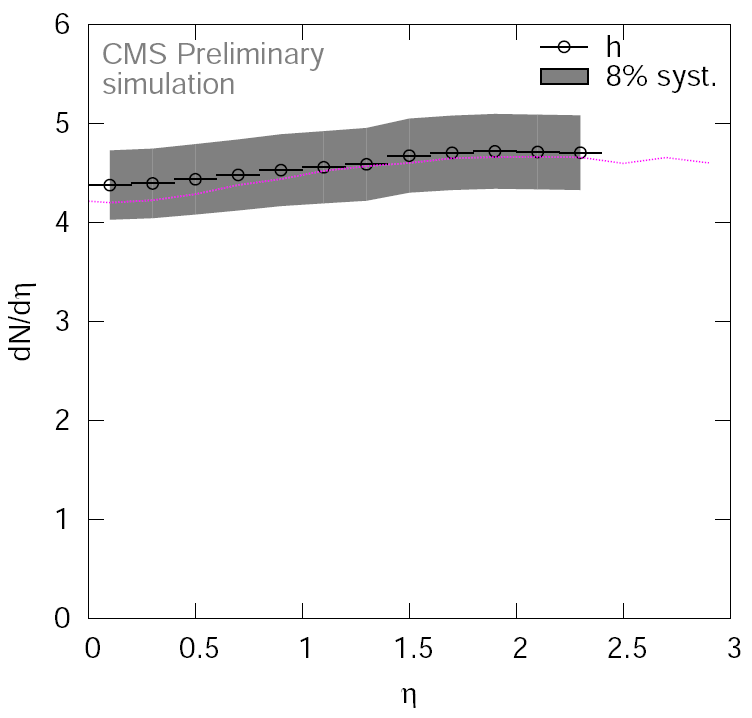}%
  \includegraphics[width=0.50\textwidth]{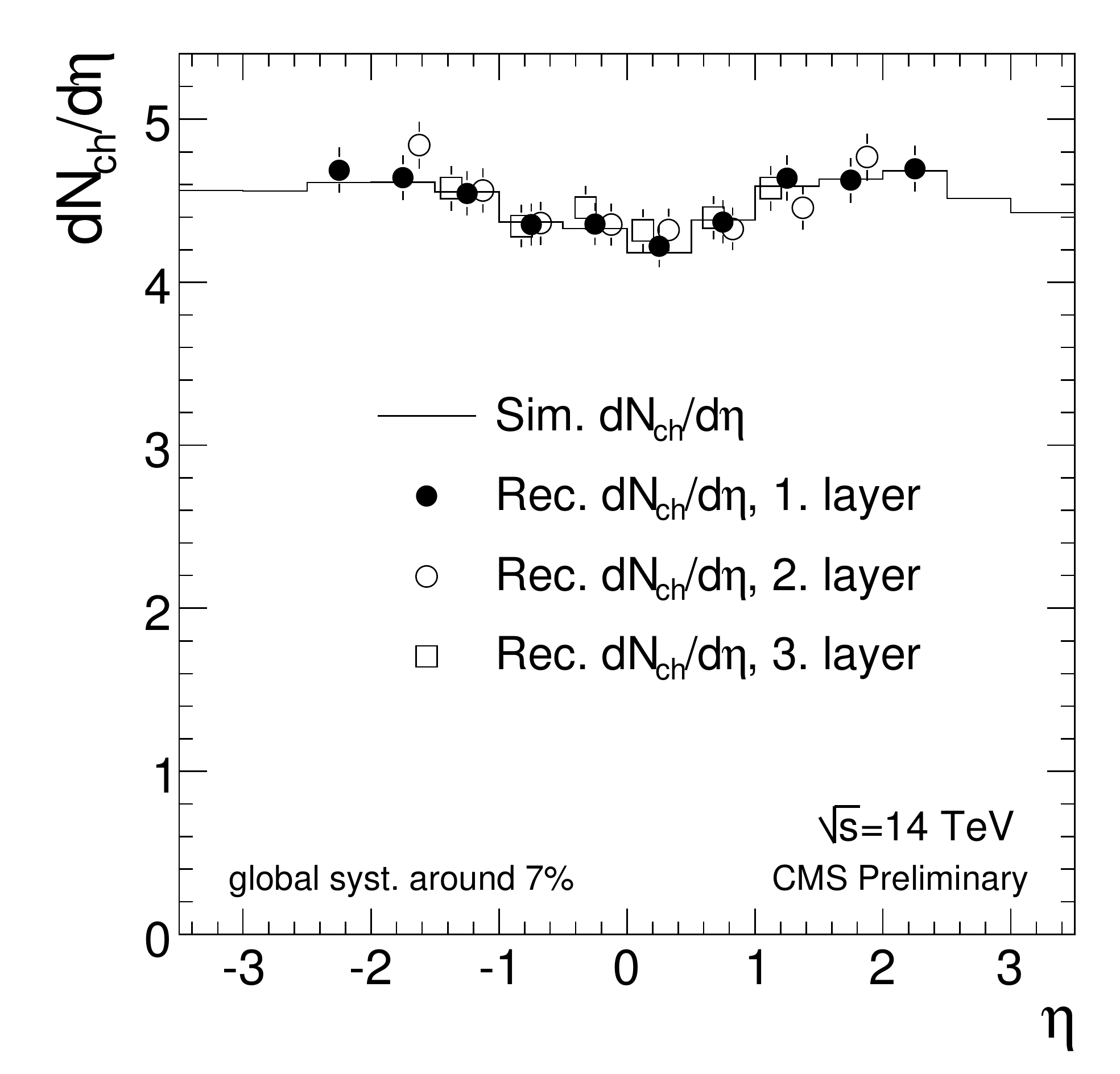}
  \caption{Pseudorapidity dependence of the charged-particle
    multiplicity at $14\TeV$ by full track
    reconstruction~\cite{CMS-PAS-QCD-07-001} as well as by applying a
    pixel hit-counting method to \PYTHIA events simulated for the CMS
    detector~\cite{CMS-PAS-QCD-08-004}.}
  \label{fig:dNdeta}
\end{figure}

Corresponding expectations by CMS for $14\GeV$ using full track
reconstruction~\cite{CMS-PAS-QCD-07-001} or a pixel hit-counting
technique~\cite{CMS-PAS-QCD-08-004}, first developed by the PHOBOS
experiment~\cite{Back:2001bq}, are shown in Figure~\ref{fig:dNdeta}.
Another method at $10\TeV$ is described by CMS
in~\cite{CMS-PAS-QCD-09-002}.

Since only a few thousand events are required for these measurements
they can be performed already early after the turn-on of a new
accelerator.  In fact, the very first measurement on LHC data, albeit
at only $900\GeV$ center-of-mass energy, has been published in the
meantime by the ALICE collaboration~\cite{Collaboration:2009dt}.
Unsurprisingly, the new results are in line with previous
measurements.  Predicted differences between $pp$ and $p\bar{p}$
scattering at the order of some permille are well below the
uncertainties and could not be observed. New publications from the LHC
experiments including also the collision data at $2.36\GeV$
center-of-mass energy should be expected soon.  In~\cite{LHCReport}
also further preliminary observations with the first LHC data have
been reported.

\newpage
Another related topic exploits the fact that the transverse region of
$60^\circ<|\Delta\phi|<120^\circ$ with respect to the leading jet in
an event is most sensitive to the Underlying Event, i.e.\ every
collision product not coming directly from the hard
scatter~\cite{Affolder:2001xt,Acosta:2004wqa}.  Extrapolations of the
UE contributions to events at LHC energies vary widely such that an
early determination of its size and the tuning of the MC generators is
an important start-up measurement.

\begin{figure}[htbp]
  \centering
  \includegraphics[width=0.45\textwidth]{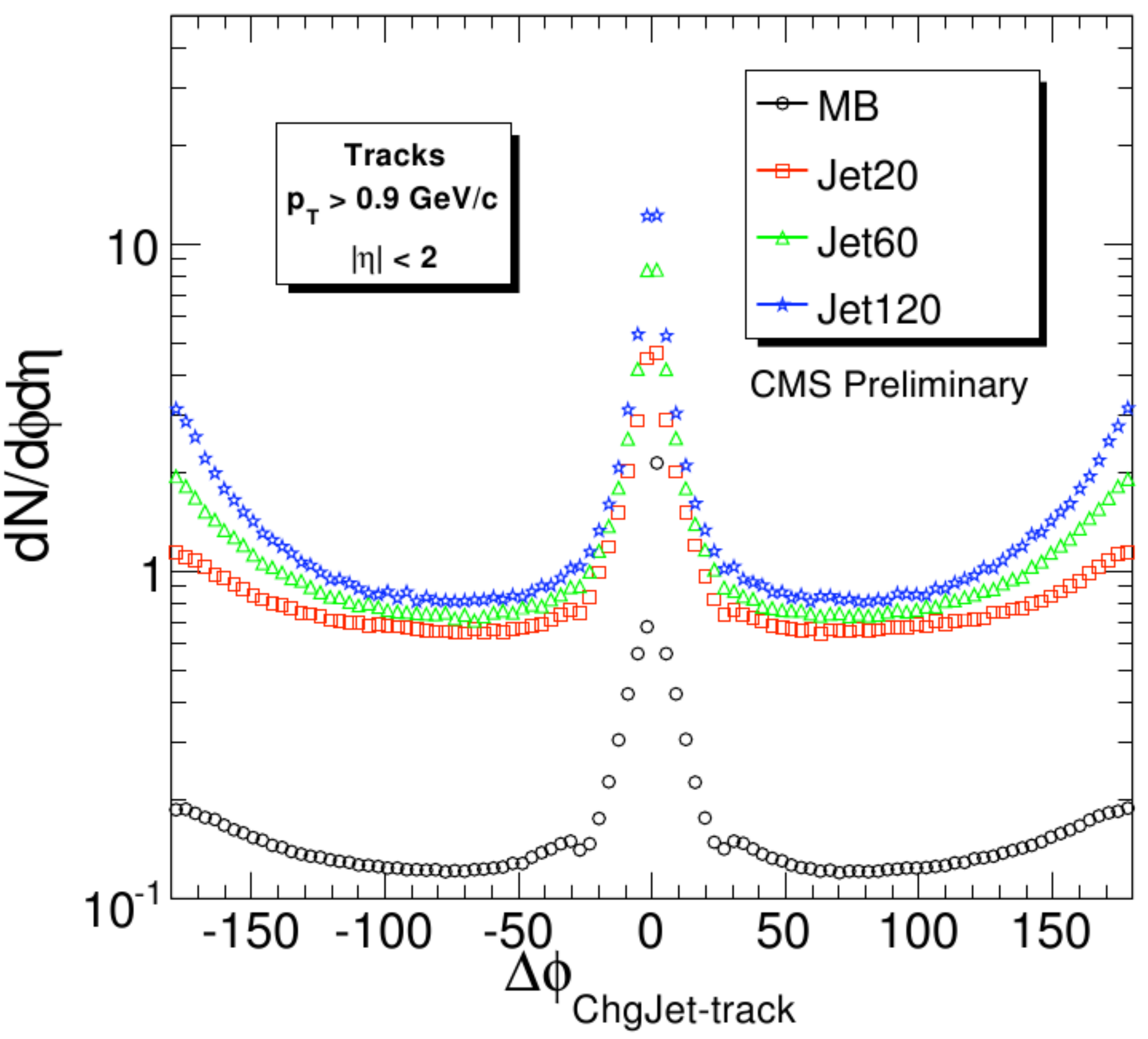}\hspace{0.08\textwidth}%
  \includegraphics[width=0.45\textwidth]{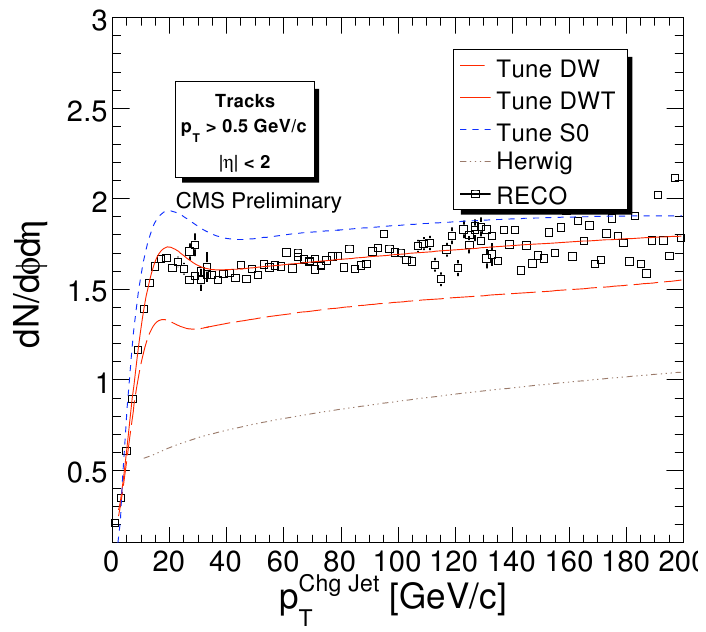}
  \caption{Composition of the total charged-particle distribution in
    $\Delta\phi$ for all trigger streams (left) and the resulting \pt
    dependence of the reconstructed charged-particle density in the
    transverse plane together with predictions of various \PYTHIA
    tunes and from \HERWIG assuming $10\pbinv$ of integrated
    luminosity at $\sqrt{s}=14\TeV$~\cite{CMS-PAS-QCD-07-003}.}
  \label{fig:UE}
\end{figure}

Figure~\ref{fig:UE} presents the composition of the total
charged-particle distribution in $\Delta\phi$ collected with Minimum
Bias and jet triggers with different jet \pt thresholds on the left
and the resulting \pt dependence of the charged-particle density in
the transverse plane on the right~\cite{CMS-PAS-QCD-07-003} as
reconstructed from CMS simulations with \PYTHIA tune DWT\@.  For
comparison the MC predictions of \PYTHIA with various tunes and from
\HERWIG without model for multiple parton interactions are shown as
well. Already with the assumed $10\pbinv$ of integrated luminosity at
$\sqrt{s}=14\TeV$ it will be possible to differentiate between the
extrapolations of some models to LHC energies. Note that in
Figure~\ref{fig:UE} right tracks with a lower limit of $\pt > 500\MeV$
have been chosen to further increase the sensitivity.

\section{Jet Measurements}

In order to establish a closer connection to the hard process which is
described theoretically in terms of partons, i.e.\ quarks, anti-quarks
and gluons, jet algorithms are employed. Although it is impossible to
unambiguously assign bunches of observed hadrons to the originating
partons, one can define a distance measure between objects and
uniquely determine which of them are sufficiently close to each other
to be considered to belong to the same jet or respectively to have a
common origin.

In total six different jet algorithms with jet sizes $R$ (or $D$)
ranging from $0.4$ to $0.7$ are in use by the ATLAS and CMS
collaborations out of which two Iterative Cone algorithms (ICone-PR,
ICone-SM, see~\cite{Cacciari:2008gp}) are not safe with respect to
comparisons with theory calculations in perturbative QCD and
therefore are not considered further. The remaining four are the
Seedless Infrared-Safe Cone algorithm (SISCone)~\cite{Salam:2007xv}
and three algorithms of the sequential recombination type: The
\kt~\cite{Ellis:1993tq,Catani:1992zp,Catani:1993hr}, the
Cambridge/Aachen~\cite{Wobisch:1998wt} and the anti-\kt
algorithm~\cite{Cacciari:2008gp}.

The standard jet measurement performed at all previous colliders so
far is the differential inclusive jet production cross
section. Unfortunately, it is affected by practically all the dominant
experimental uncertainties due to the jet energy calibration (JEC),
the luminosity determination, the jet energy resolution (JER), trigger
efficiencies, and, less important, the spatial resolutions in
azimuthal angle and pseudorapidity.

\begin{figure}[htbp]
  \centering
  \includegraphics[width=0.50\textwidth]{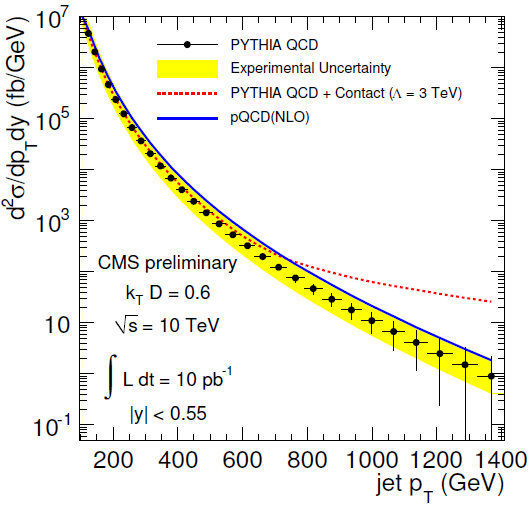}%
  \includegraphics[width=0.50\textwidth]{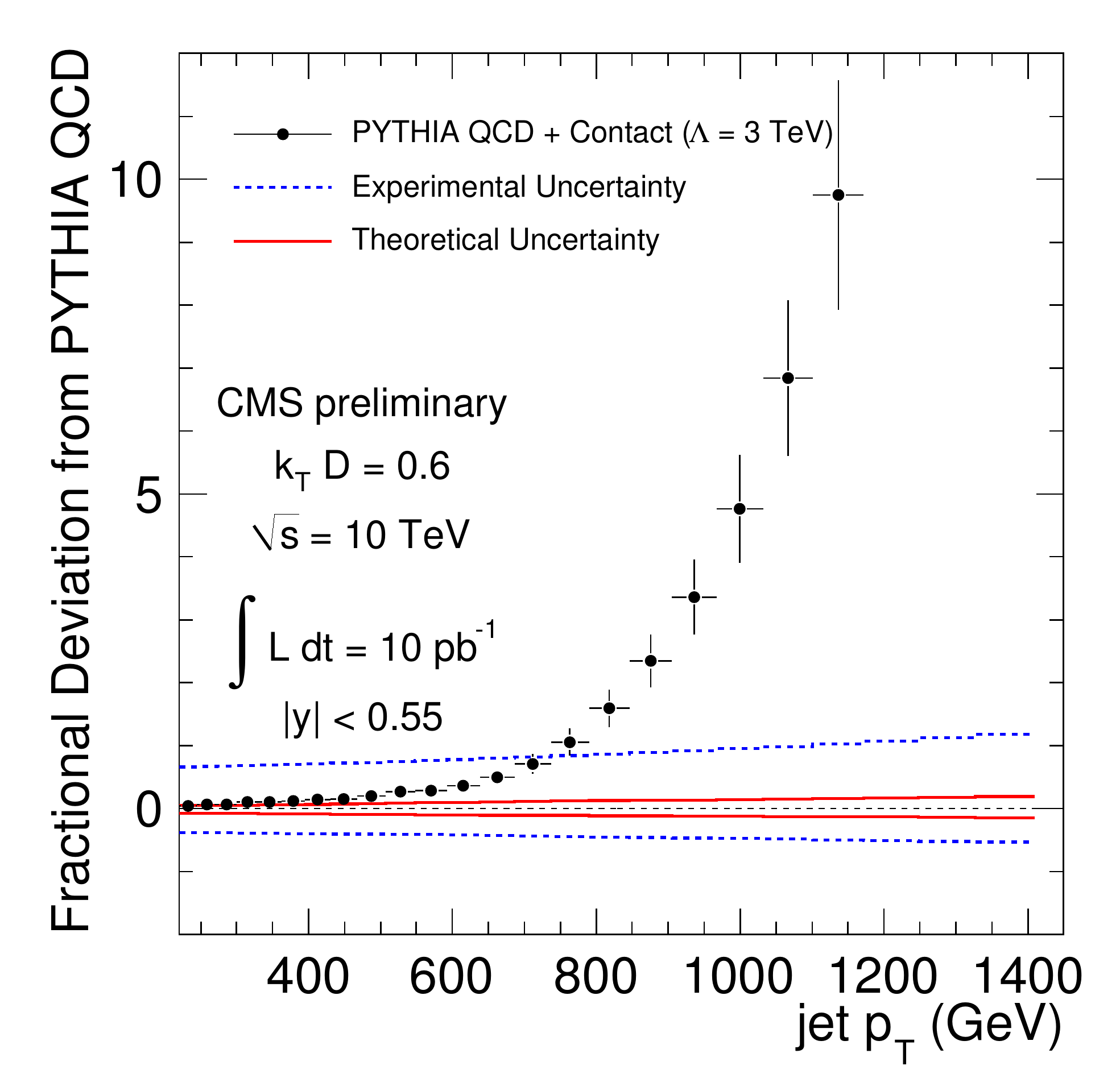}
  \caption{Measured inclusive jet spectrum (K factors times \PYTHIA
    with CMS simulation) with experimental systematic uncertainty
    compared with theory (NLO times non-perturbative corrections) and
    \PYTHIA QCD+3\TeV contact interaction term (left). Fractional
    difference of the QCD+contact interaction term and pure \PYTHIA
    QCD is shown in comparison to the experimental and theoretical
    uncertainties (right).}
  \label{fig:InclJetsComparison}
\end{figure}

The reach in jet transverse momentum, however, is beyond any previous
collider experiment~\cite{Aaltonen:2008eq, Abulencia:2007ez,
  Abazov:2008hu} already with $10\pbinv$ of integrated luminosity at a
center-of-mass energy of $\sqrt{s}=10\TeV$.  Indications of new
physics like from contact interactions would clearly be observable as
demonstrated by CMS in~\cite{CMS-PAS-QCD-08-001}.
Figure~\ref{fig:InclJetsComparison} compares the inclusive jet cross
section in such a contact interaction scenario for a compositeness
scale of $\Lambda^+ = 3\TeV$ with the pure QCD prediction including
estimates of all relevant experimental and theoretical uncertainties.

Examples for jet observables less sensitive to experimental
uncertainties are angle-related and/or normalized like dijet azimuthal
decorrelations and event shapes or cross-section ratios like the dijet
production ratio in pseudorapidity and 3-jet to all-jet ratios.  Here,
the luminosity uncertainty is eliminated and the uncertainty due to
the JEC is largely reduced. Expectations for LHC with this type of
analyses can be found for example in the
references~\cite{CMS-PAS-QCD-09-003,CMS-PAS-QCD-08-003,CMS-PAS-SBM-07-001}.


\section[Weak Boson Cross-Sections, $W$ Mass]{Weak Boson Cross-Sections, \boldmath$W$\unboldmath Mass}

In contrast to the previously presented reactions the production rates
for the weak bosons are orders of magnitudes smaller than for Minimum
Bias or jet events (depending on \pt). Nevertheless approximate rates
of $10/$s resp.\ $3/$s for $W$ resp.\ $Z$ bosons and theoretical
uncertainties smaller than $1$\% allow for precision measurements that
stringently test the Standard Model in a new energy regime.
Experimental uncertainties are well under control as long as the
leptonic decay modes into muons and electrons are concerned.  For
selections of isolated leptons plus missing transverse momentum or
unlike-sign lepton pairs within $|\eta_l| < 2.5$ and for $\pt >
15$--$25\GeV$ an accuracy of the $W$ and $Z$ production cross sections
of about $5\%$ for $W$'s and $3\%$ for $Z$'s is expected by ATLAS for an
integrated luminosity of $50\pbinv$ at $14\TeV$~\cite{Aad:2009wy}.
With $1\fbinv$ of data irreducible uncertainties of about $1$--$2$\%
are anticipated. Corresponding estimations by CMS can be found
in~\cite{CMS-PAS-EWK-09-001,CMS-PAS-EWK-09-004}.

\begin{figure}[htbp]
  \centering
  \includegraphics[width=0.51\textwidth]{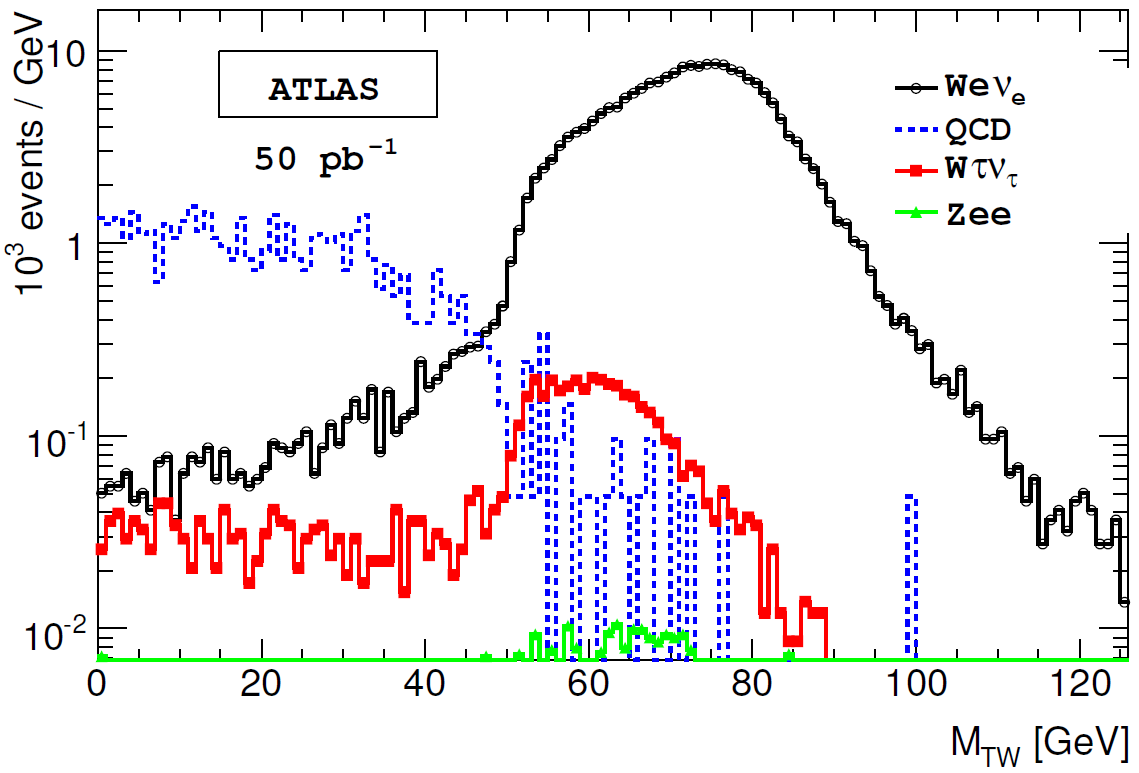}%
  \includegraphics[width=0.49\textwidth]{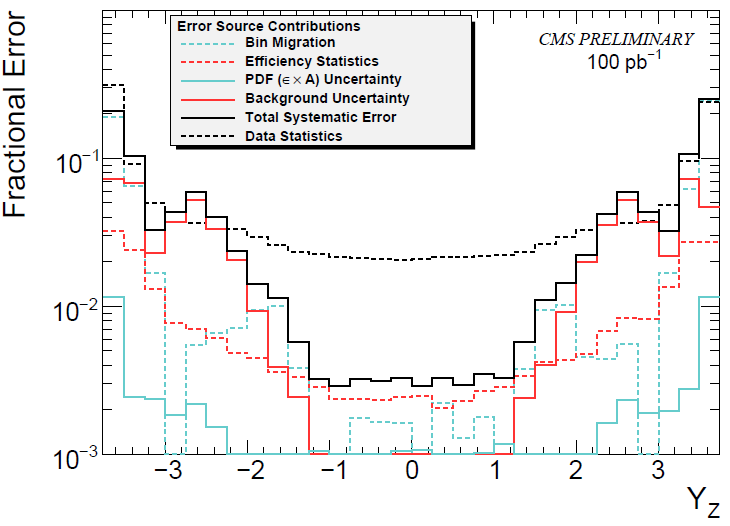}
  \caption{$W$ transverse mass distribution in the $W\rightarrow
    e\nu_e$ channel (left)~\cite{Aad:2009wy} and fractional
    uncertainties of a $Z$ rapidity measurement
    (right)~\cite{CMS-PAS-EWK-09-005}.}
  \label{fig:MWtransZdiff}
\end{figure}

Improving on the $W$ mass is more difficult, especially considering
the reduced uncertainty of $\Delta M_W = 31\MeV$ from CDF and D0
presented at this conference~\cite{Wallny:RADCOR2009}.  Compared to
the PDG Review 2009~\cite{Amsler:2008zzb} with $\Delta M_W = 40\MeV$
from Tevatron and $25\MeV$ as world combination this is a remarkable
progress.  On a longer timescale ATLAS predicts an achievable
precision of $\order(< 10\MeV)$ for $10\fbinv$ of integrated
luminosity at $14\TeV$ provided radiative corrections are under
control on the theory side~\cite{SN-ATLAS-2008-070,Besson:2008zs}. An
example for a $W$ transverse mass distribution in the $W\rightarrow
e\nu_e$ channel for $50\pbinv$ at $14\TeV$ is displayed in
Figure~\ref{fig:MWtransZdiff} left.

\section{Differential Weak Boson Measurements}

\begin{figure}[htbp]
  \centering
  \includegraphics[width=0.40\textwidth]{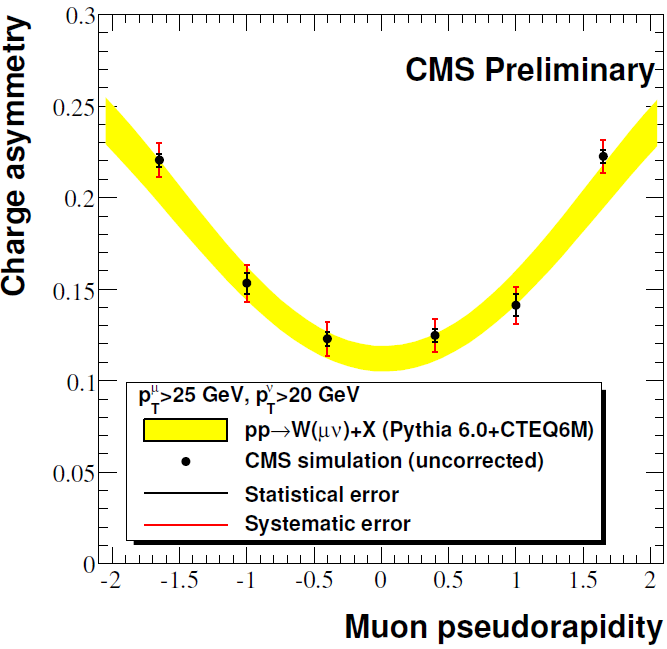}%
  \raisebox{1em}{\includegraphics[width=0.60\textwidth]{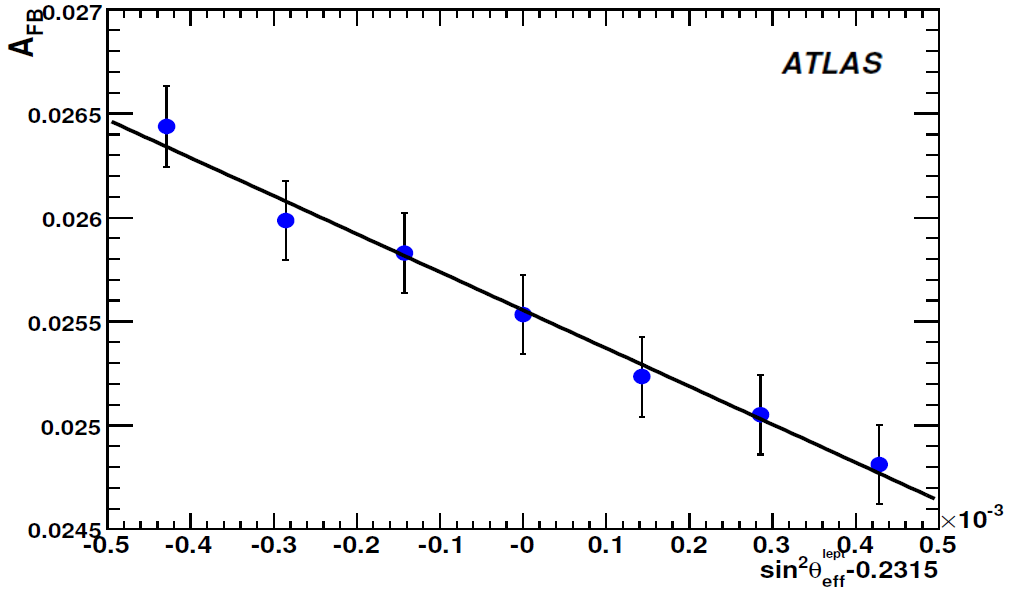}}
  \caption{The reconstructed $W$ charge asymmetry including estimated
    statistical and systematic uncertainties for $100\pbinv$ of
    simulated luminosity at $10\TeV$ from CMS
    (left)~\cite{CMS-PAS-EWK-09-003} and the forward backward
    asymmetry $A_{FB}$ versus the weak mixing angle
    $\sin^{2}\theta^{\rm lept}_{\rm eff}$ at the $Z$ pole for
    $100\fbinv$ of integrated luminosity at $14\GeV$ from ATLAS
    (right)~\cite{Aad:2009wy}.}
  \label{fig:WchasymFBasym}
\end{figure}

With somewhat more integrated luminosity the differential
distributions of the weak bosons, in particular for the $Z$, can be
exploited where the $Z$ transverse momentum provides even more
constraints on QCD, especially on non-perturbative effects of initial
parton emissions, while the rapidity distribution directly probes the
parton density functions (PDFs) of the proton. Estimates on the
fractional uncertainties for the $Z$ rapidity distribution from
CMS~\cite{CMS-PAS-EWK-09-005} are displayed in
Figure~\ref{fig:MWtransZdiff} right. Equally, the $W$ charge asymmetry
shown in Figure~\ref{fig:WchasymFBasym} left can start constraining
the PDFs with only $50\pbinv$ of data~\cite{CMS-PAS-EWK-09-003}.  On a
much longer timescale, once, about $100\fbinv$ of integrated
luminosity at highest LHC energies have been accumulated, the weak
mixing angle comes in reach for improving its accuracy. This has been
studied by ATLAS in~\cite{Aad:2009wy}. The lever arm for this is
pictured in Figure~\ref{fig:WchasymFBasym} right.

\section{Boson plus Jet and Di-Boson Production}

\begin{figure}[htbp]
  \centering
  \raisebox{0.5em}{\includegraphics[width=0.455\textwidth]{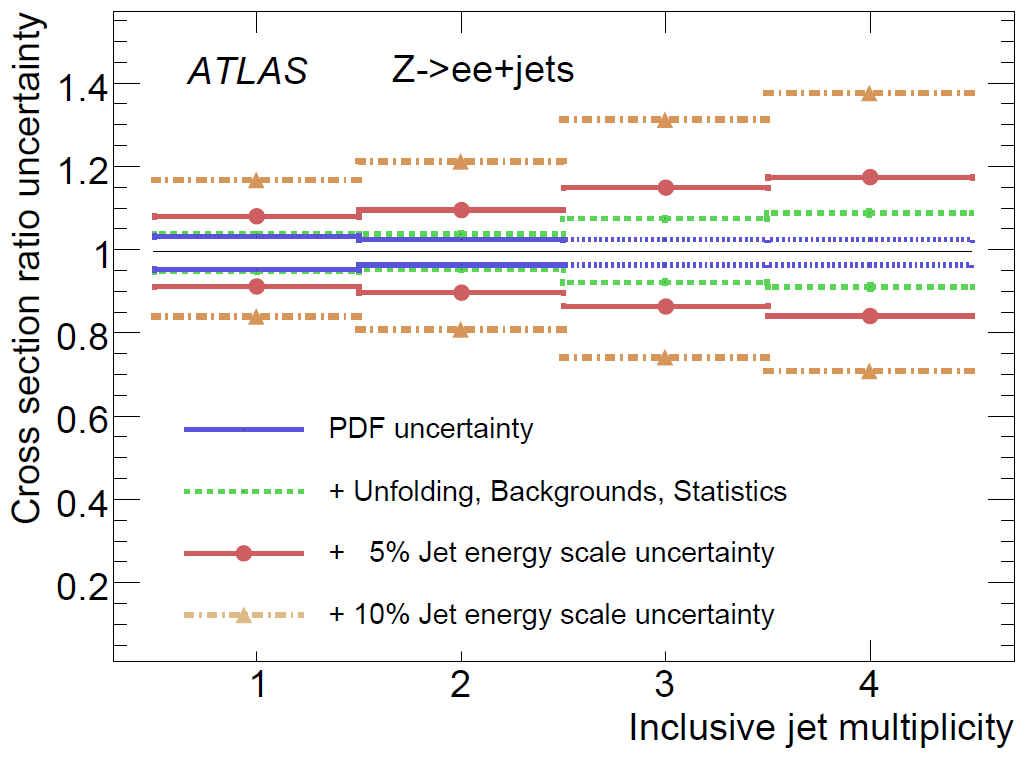}}%
  \includegraphics[width=0.545\textwidth]{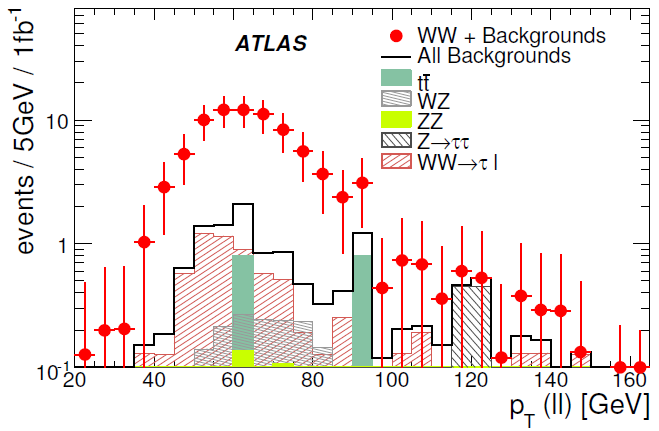}
  \caption{Relative uncertainties on a data-theory comparison of the
    $Z+$jet multiplicity cross sections in the $Z\rightarrow ee$
    channel (left)~\cite{ATLASPublicWeb} and the \pt distribution of
    lepton pairs for simulated $WW$ candidate events
    (right)~\cite{Aad:2009wy} both for $1\fbinv$ of integrated
    luminosity at $14\TeV$.}
  \label{fig:ZjetWW}
\end{figure}

Finally, the much higher center-of-mass energy at the LHC allows for
more precise studies than ever before of multiple boson or of boson
plus jet production. Figure~\ref{fig:ZjetWW} shows an estimate on the
fractional uncertainties for the $Z$+jet cross sections in the
$Z\rightarrow ee$ channel (left)~\cite{ATLASPublicWeb} as well as the
\pt distribution of candidate lepton pairs for $WW$ di-boson events
together with backgrounds (right)~\cite{Aad:2009wy} both from ATLAS
for $1\fbinv$ of integrated luminosity at $14\TeV$. A study by CMS on
$Z+$jet production can be found in~\cite{CMS-PAS-EWK-08-006}. The CMS
potential for measuring $WW$ production with $100\pbinv$ at $10\TeV$
is reported in~\cite{CMS-PAS-EWK-09-002}.

\newpage
\section{Outlook}

The first LHC collision data have been registered up to a
center-of-mass energy of $2.36\TeV$ and the LHC experiments are in
full swing of commissioning their detectors and publishing first
physics results. Even higher energies will be reached in the very near
future opening up a window to an unprecedented multitude of physics
analyses involving multiple jet and boson production studies that were
not possible before. The rich program of new physics measurements not
only re-establishes the Standard Model but also sets the scene for
searches for new phenomena. This year marks the beginning of a new era
of particle physics.

\acknowledgments

I would like to thank the organizers for the kind invitation to
participate in this conference.

%
%
\bibliographystyle{JHEP_mine}
\bibliography{my-references,cms-pub,lhc,qcd}

\end{document}